\documentstyle[11pt,newpasp,epsf,psfig]{article}

\begin{document}

\title{Gas and stellar 2D kinematics in early-type galaxies}
\author{Eric Emsellem}
\affil{Centre de Recherche Astronomique de Lyon, 9 av, Charles Andr\'e, 69561 Saint-Genis Laval Cedex, France}

\author{Paul Goudfrooij}
\affil{Space Telescope Science Institute, 3700 San Martin Drive, Baltimore, MD 21218, USA,
Affiliated to the Astrophysics Division, SSD, European Space Agency}
\begin{abstract}
We have obtained integral field spectroscopy of a small sample of early-type
galaxies to study the kinematical coupling between the stellar and gaseous components
in their central regions. 
\end{abstract}


\section{Introduction}

It is surprising to see how widely accepted is the assumption that
giant early-type galaxies tend to have triaxial
shapes (fainter ones being predominantly axisymmetric), when
strong and convincing cases of triaxiality are rare 
(e.g. Merritt 1999 and references therein).
Fits to the kinematics of luminous early type galaxies
using axisymmetric models are surprisingly good, although this may be linked, 
in most cases (but see Statler, Dejonghe, \& Smecker-Hane 1999), to a critical lack of detailed 
kinematical information. Thus a more accurate statement would be:
we still do not know much about the detailed intrinsic shape and dynamics
of early-type galaxies.

In this context, two-dimensional kinematical maps are a prerequisite
for the determination of the underlying gravitational potential.
Stars contribute for most of the visible mass of early-type galaxies
and their motions are (almost exclusively) determined by gravitation.
However, general axisymmetric and triaxial dynamical models are not easy 
to build, mainly because of the very large solution space to probe. 
Gas orbits are thought to be simpler to deal with (as generally
assumed circular or elliptical), but non-gravitational motions
can enter the play, particularly in the central regions.
The realisation, not that long ago (see e.g. Goudfrooij 1997 
and references therein), that most early-type galaxies do 
contain a significant gaseous component, led us to start a program
to obtain the 2D kinematics of the stellar AND gaseous components in the central
regions of a small sample of early-type galaxies.

\section{Observations}
We have observed about a dozen early-type galaxies using the {\tt TIGER} Integral Field Spectrograph 
(IFS) at the CFH Telescope. The {\tt TIGER} spectrograph 
provided about 400 spatial elements, homogeneously covering the field of view
with a spatial sampling of $0\farcs39$.

To obtain both the stellar and gas kinematics, we observed
two spectral domains, namely a {\tt blue} domain around 5200\AA, including
the Mg triplet as well as Ca and Fe stellar absorption lines, and a {\tt red} one
including the H$\alpha$, [NII] and [SII] emission lines.
The spectral sampling was 1.5\AA\ per pixel, with a final resolution
of 1700 and 2200 in the {\tt blue} and {\tt red}, respectively.

\begin{figure}[h]
\centerline{\psfig{figure=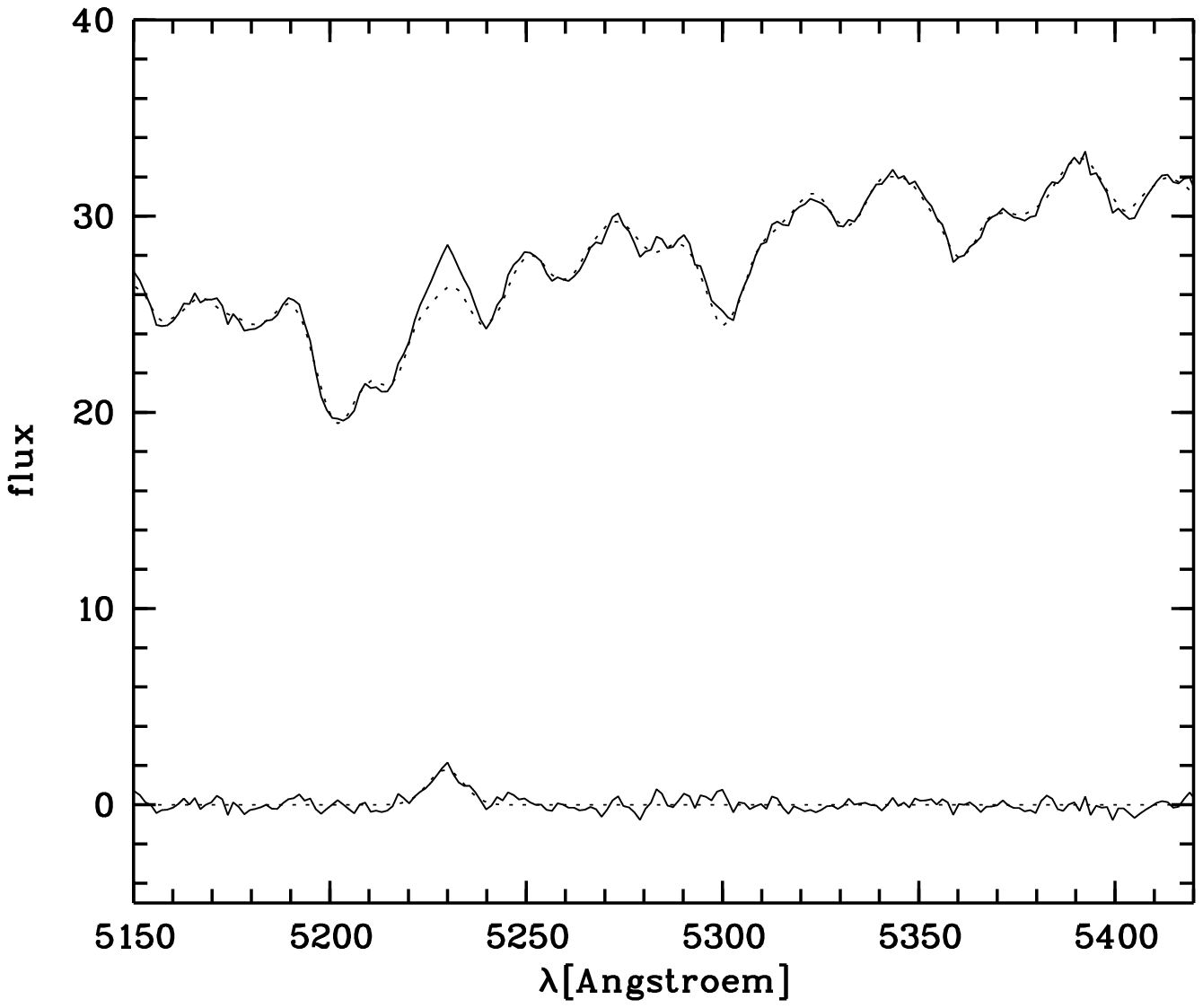,width=6.5cm}
\psfig{figure=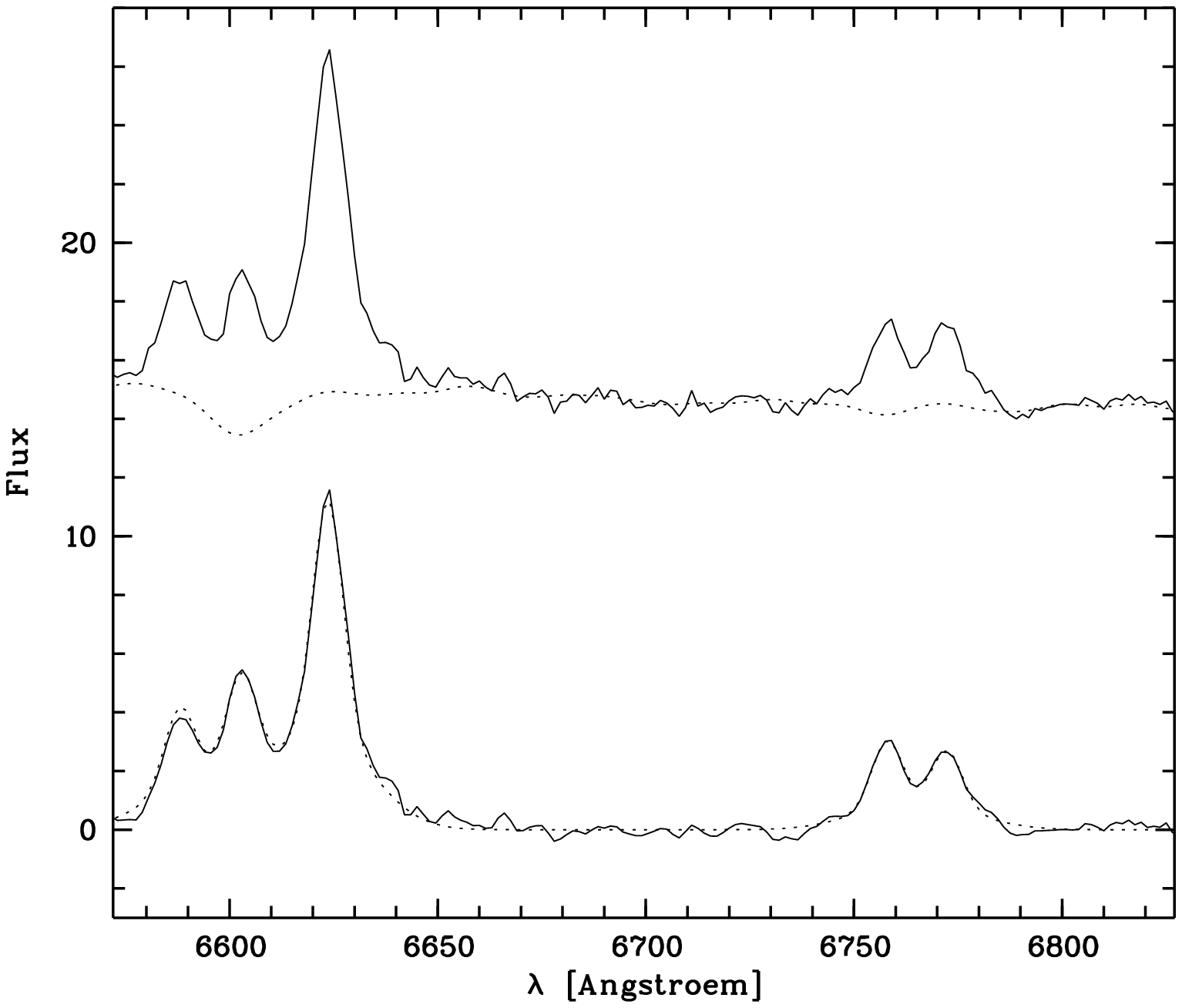,width=6.5cm}}
\caption{Example of stellar continuum subtraction for a {\tt blue} (left) and {\tt red}
(right) spectra of NGC~2974. Top: original spectra (solid) and their best fit continuum spectra
(dotted). Bottom: the continuum subtracted spectra (solid) and the emission line fits
(dotted).}
\label{fig:NII}
\end{figure}

The data have been reduced using a dedicated software developed
at the Lyon Observatory (Rousset, PhD Thesis, Lyon). 
The two major difficulties were: first to 
correct the {\tt blue} spectra from
the contamination by the [NI]$\lambda$5200 emission line (when present), and second
to properly subtract the stellar contribution (mainly the H$\alpha$ absorption 
line) from the {\tt red} spectra. This was achieved with an algorithm 
which includes a library of stellar and galaxy spectra (coll. Paul Goudfrooij). 
Illustrative examples of the resulting subtractions are given in Fig.~\ref{fig:NII}.
Maps of the distribution and kinematics of the 
gas and stellar components were then built for all the galaxies in the sample,
and will soon be published in a forthcoming paper through a collaboration
with P. Goudfrooij (StSci) and P. Ferruit (Uni. of Maryland \& CRA Lyon).

\section{Results}
\begin{figure}
\centerline{\psfig{figure=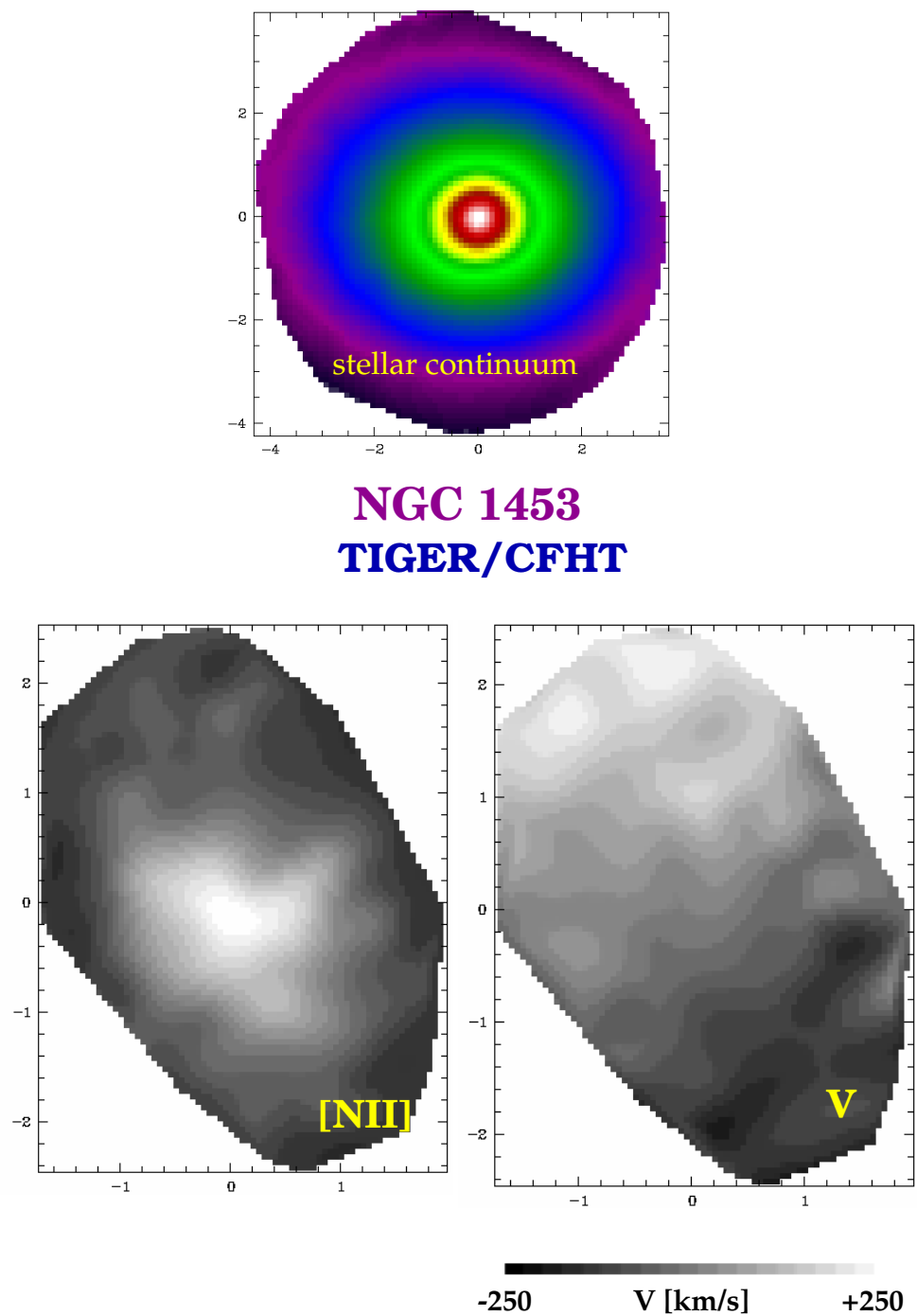,width=13cm}}
\caption{{\tt TIGER} data of NGC~1453: stellar continuum (V band) reconstructed
image (top), the ionised gas distribution map (bottom left, [NII]$\lambda$6584) 
and the corresponding gas velocity field (bottom right).}
\label{fig:n1453}
\end{figure}
The ionised gas distribution in these galaxies exhibit a variety 
of morphologies, including nuclear spirals (e.g. NGC~2974, NGC~4278),
point-like emission regions (NGC~3414, NGC~6482), corotating disks (NGC~5838, NGC~2749)
and counter-rotating discs (NGC~128).
In some objects, part of the gas is clearly coupled with the dust component
(e.g. NGC~4374, see Bower et al. 1997), although this is not a systematic feature.
In most cases, the gas and stars have different angular momentum axis.
One striking example is NGC~1453, in which they are tilted by about 50 degrees 
with respect to each other (Fig.~\ref{fig:n1453}), suggesting a triaxial
geometry (see also Pizzella et al. 1997).

The blended H$\alpha$/[NII] emission line system often exhibits 
broad wings which could be interpreted as resulting from the presence of a
broad H$\alpha$ line (a BLR). However, in all cases, these broad wings are also observed in the
forbidden [SII] emission lines, although with a lower contrast.
This argues for an unresolved kinematical gradient in the centre
of these galaxies. This is confirmed by the spatial mapping of these wings,
whose presence is limited to an unresolved central peak.
The fact that these wings are weaker in the [SII] lines
could be naturally explained by its lower critical density,
which diminishes the contribution of high-density regions.
This obviously does not mean that BLRs are not present,
but set an upper limit on their contribution
to the central emission line spectra.

All our spectra are compatible with the LINER type, although
at different levels of activity. 
This is consistent with the compilation done by Ho, Fillipenko, \& Sargent (1997)
for the 7 objects in common. It is interesting to note
that two of the four most active galaxies in our sample 
do show the presence of a nuclear spiral, the third one (M~87) having
a spiral-like gas disc, and the fourth being viewed nearly edge-on.
Central spiral structures may therefore play a role in the nuclear activity
(see also Regan \& Mulchaey 1999).

\section{Perspectives}
While our sample is not complete in any sense, it is striking
to observe the morphological and kinematical decoupling of the ionised gas 
with respect to the stellar component. This certainly hints for an external
origin in most cases. We plan to continue this study by using the recently
comissioned integral field spectrograph OASIS, mounted on the adaptive optics bonnette
of the CFHT. Our understanding of the gas/stars coupling in early-type
galaxies will also greatly benefit from the on-going survey at the WHT 
conducted by the {\tt SAURON} Consortium (Lyon/Leiden/Durham).

\end{document}